%% file: main.tex
\documentclass[runningheads]{llncs}
\usepackage{graphicx}
\usepackage{subfigure}

\usepackage{multicol, latexsym, amsmath, amssymb}
\usepackage{blindtext}
\usepackage{caption}

\usepackage{csquotes}
\usepackage{alltt}
\usepackage{listings}
\usepackage{amsmath}
\usepackage{pgfplots}
\usetikzlibrary{patterns}
\usepackage{xcolor}
\usepackage{soul}

\usepackage{float}
\floatstyle{ruled}
\newfloat{program}{t}{lop}
\floatname{program}{Program}
\usepackage{array}

\pgfdeclarepatternformonly{mynewdots}{\pgfqpoint{-1pt}{-1pt}}{\pgfqpoint{1pt}{1pt}}{\pgfqpoint{1pt}{1pt}}
{%
  \pgfpathcircle{\pgfqpoint{0pt}{0pt}}{.5pt}%
  \pgfusepath{fill}%
}%

\begin{document}
\title{Template-Based Conjecturing for Automated Induction in Isabelle/HOL}
%
%
\author{Yutaka Nagashima\orcidID{0000-0001-6693-5325} \and
Zijin Xu\orcidID{0000-0001-7230-0131} \and
Ningli Wang\orcidID{0000-0001-7775-436X} \and
Daniel Sebastian Goc\orcidID{0000-0002-2347-8037} \and
James Bang\orcidID{0000-0001-9345-3479}}
%
\authorrunning{Y. Nagashima et al.}
%
\institute{Huawei Cambridge Research Centre}
%
\maketitle              
\begin{abstract}
Proof by induction plays a central role in formal verification. 
However, its automation remains as a formidable challenge in Computer Science.
To solve inductive problems, human engineers often have to provide auxiliary lemmas manually.
We automate this laborious process with \textit{template-based conjecturing}, 
a novel approach to generate auxiliary lemmas and use them to prove final goals.
Our evaluation shows
that our working prototype, TBC, achieved
40 percentage point improvement of success rates for problems at intermediate difficulty level.

\end{abstract}
\section{Introduction}\label{introduction}

Consider the following definitions of \verb|add| and \verb|even| on natural numbers:

\begin{alltt}
add      0  \(m\) = \(m\)
add (Suc \(n\)) \(m\) = Suc (add \(n\) \(m\))

even           0   = True
even      (Suc 0)  = False
even (Suc (Suc \(n\))) = even \(n\)
\end{alltt}

\noindent
Intuitively, the following statement holds: \texttt{even (add \(n\) \(n\))}.

However, if we apply structural induction on \textit{n}, the simplification 
based on the definitions of \verb|add| and \verb|even| gets stuck at
\texttt{even (add \(n\) \(n\))} $\Rightarrow{}$ \texttt{even (S (add \(n\) (S \(n\))))}
when attacking the induction step.
This is due to the definition of \texttt{add}, which does not allow us to operate on its second argument.
Hence, if we want to prove this statement, we need to introduce auxiliary lemmas.



What lemmas should we introduce?
Empirically, we know various mathematical structures share well-known \textit{algebraic properties} such as associativity and commutativity. 
For example, our example problem uses \texttt{add}, 
which satisfies the following properties:

\begin{alltt}
add \(n\) (add \(m\) \(k\)) = add (add \(n\) \(m\)) \(k\) \hfill (add is associative)
add \(n\) \(m\) = add \(m\) \(n\)  \hfill (add is commutative)
\end{alltt}

The commutative property of \texttt{add} 
allows us to operate on its second argument. 
Hence, if we prove this property, 
we can revert back to the original goal and finish its proof.

To automate this process, this paper introduces TBC, a tool that 
produces such \underline{t}emplate-\underline{b}ased \underline{c}onjectures
and attempts to prove them as well as the original proof goal in Isabelle/HOL \cite{isabelle}.
For example, when applied to \texttt{even (add \(n\) \(n\))},
TBC first proves 10 conjectures then proves the original goal using two of them as shown in 
Program \ref{code:generated_proof} in Appendix.

We chose Isabelle/HOL to exploit its powerful proof tactics and
counter-example finders;
however, the underlying idea of template-based conjecturing is not
specific to Isabelle/HOL:
we can build similar systems for other provers
if they are equipped with equivalent tools.
We developed TBC under the following research hypothesis:
\begin{quote}
We can improve the proof automation of inductive problems
by producing and proving conjectures based on fixed but general properties
about relevant functions.
\end{quote}

\noindent
Our contributions are:
\begin{itemize}
    \item the working prototype of a powerful inductive prover based on template-based conjecturing and newly developed default strategy (Section \ref{sec:overview}),
    \item the identification of useful properties (Section \ref{sec:properties}), and
    \item extensive evaluations of TBC to test our research hypothesis (Section \ref{sec:evaluation}).
\end{itemize}

\section{System Description}

\begin{figure}[ht!]
\centering
\includegraphics[width=0.7\textwidth]{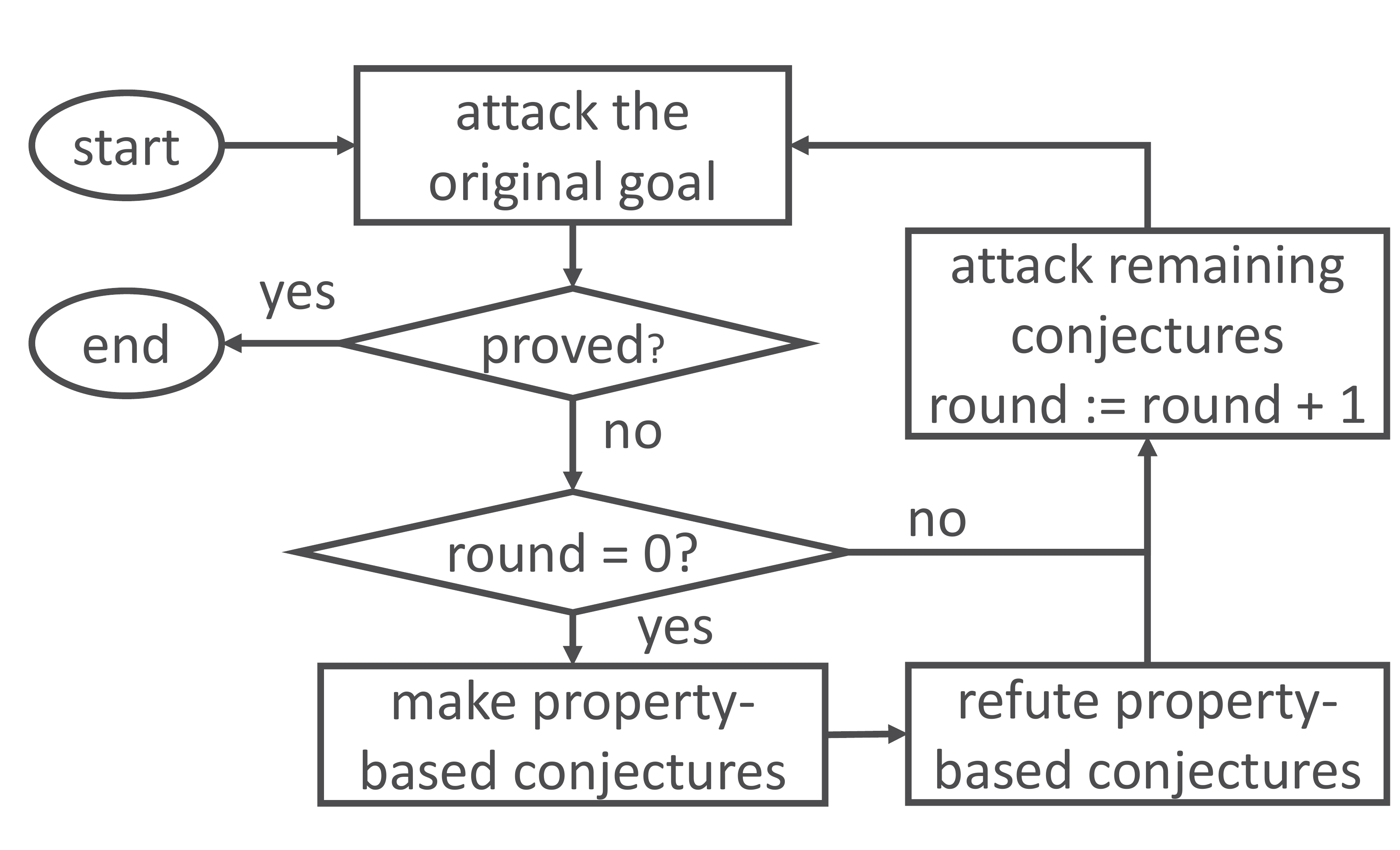}
\caption{Workflow of TBC} \label{workflow}
\end{figure}

\subsection{Overview}\label{sec:overview}
Figure \ref{workflow} shows how TBC attacks inductive problems using template-based conjecturing.
Given an induction problem, the tool first attempts to prove the goal 
using a default strategy, \verb|TBC_Strategy|, written in the proof strategy language, PSL \cite{psl}.
As shown in Program \ref{program:default_strategy},
\verb|TBC_Strategy| combines Isabelle's proof tactics, such as \verb|auto| and \verb|clarsimp|, 
and other sub-tools, such as smart induction \cite{lifter,smart_induct,sem_ind} and Sledgehammer \cite{sledgehammer} to prove the goal completely.
That is, PSL uses Sledgehammer as a sub-tool,
even though Sledgehammer itself is a meta-tool that uses external provers and Isabelle's tactics
to prove given problems.

As PSL is a new meta-tool,
we first explain the language constructs in Program \ref{program:default_strategy}.
\texttt{Ors} is a combinator for deterministic choice, whereas
\texttt{Thens} and \texttt{PThenOne} combine sub-strategies sequentially.
\texttt{Subgoal} lets PSL focus on the first sub-goal, 
temporarily hiding other sub-goals from the scope,
while \texttt{IsSolved} checks if all proof obligations are solved within the current scope.
\texttt{Auto}, \texttt{Clarsimp}, and \texttt{Fastforce} correspond to Isabelle's default
tactics of the same name, while \texttt{Hammer} calls Sledgehammer \cite{sledgehammer}
and \texttt{Smart\_Induct} applies 5 promising candidates of proof by induction \cite{smart_induct,sem_ind}.
Essentially, this strategy applies increasingly expensive sub-strategies to solve
proof goals using backtracking search.

If \verb|TBC_Strategy| fails to prove the goal, 
it produces conjectures based on properties specified in advance,
following the process explained in Section \ref{sec:properties}.
Then, the tool attempts to refute the conjectures using Isabelle's counter-example generators: 
Quickcheck \cite{quickcheck} and Nitpick \cite{nitpick}.
After filtering out refuted conjectures,
TBC attempts to prove the remaining conjectures using the default strategy.
While doing so, TBC registers proved conjectures as auxiliary lemmas, 
so that it can use them to prove other conjectures.

\begin{program}[ht!]
\begin{alltt}
Ors [
  Thens [Auto, IsSolved],
  PThenOne [Smart_Induct, Thens [Auto, IsSolved]],
  Thens [Hammer, IsSolved],
  PThenOne [
    \hl{Smart_Induct},
    Ors
      [Thens [
         Repeat (
           Ors [
             Fastforce,
             \hl{Hammer},
             Thens [ Clarsimp, IsSolved ],
             Thens [
               \hl{Subgoal},
               \hl{Clarsimp},
               Repeat (
                 Thens [ \hl{Subgoal},
                         Ors [ Thens [Auto, IsSolved],
                               Thens [ \hl{Smart_Induct}, \hl{Auto}, \hl{IsSolved} ] ] ]
               ),
               \hl{IsSolved}
             ]
           ]
         ),
         \hl{IsSolved}   
       ]
    ]
  ]
]
\end{alltt}
\caption{\texttt{TBC\_Strategy}: TBC's default strategy.}
\label{program:default_strategy}
\end{program}

For example, \texttt{TBC\_Strategy} finds the following proof script for
the commutativity of \verb|add|.
To demonstrate how \texttt{TBC\_Strategy} finds proofs using backtracking search,
we highlighted the parts of Program \ref{program:default_strategy} 
that were \textit{not} backtracked but resulted in this script.
We invite readers to compare these highlighted parts in Program \ref{program:default_strategy}  against the resulting
script and to find out which proof tactic Sledgehammer used to prove the corresponding sub-goal. \footnote{Answer: Sledgehammer used the \texttt{simp} tactic with an auxiliary lemma about identity. Furthermore, \texttt{IsSolved} resulted in the \texttt{done} command in the script.}

\begin{alltt}
lemma commutativity: "add \(var_1\) \(var_2\) = add \(var_2\) \(var_1\)"
  apply ( induct_tac "\(var_1\)" )
    apply ( simp add : identity )
    subgoal
      apply clarsimp
      subgoal
        apply ( induct_tac "\(var_2\)" )
        apply auto
      done 
    done 
  done
\end{alltt}

After processing the list of conjectures,
TBC comes back to the original goal.
This time, it attacks the goal,
using proved conjectures as auxiliary lemmas.
If TBC still fails to prove the original goal,
it again attacks the remaining conjectures hoping that proved conjectures may help the strategy to prove remaining ones.
By default, TBC gives up after the second round
and shows proved conjectures and their proofs in Isabelle's standard editor's 
output pane,
so that users can exploit them when attacking original goals manually.

\begin{figure}[hp!]
\centering
\includegraphics[width=\textwidth]{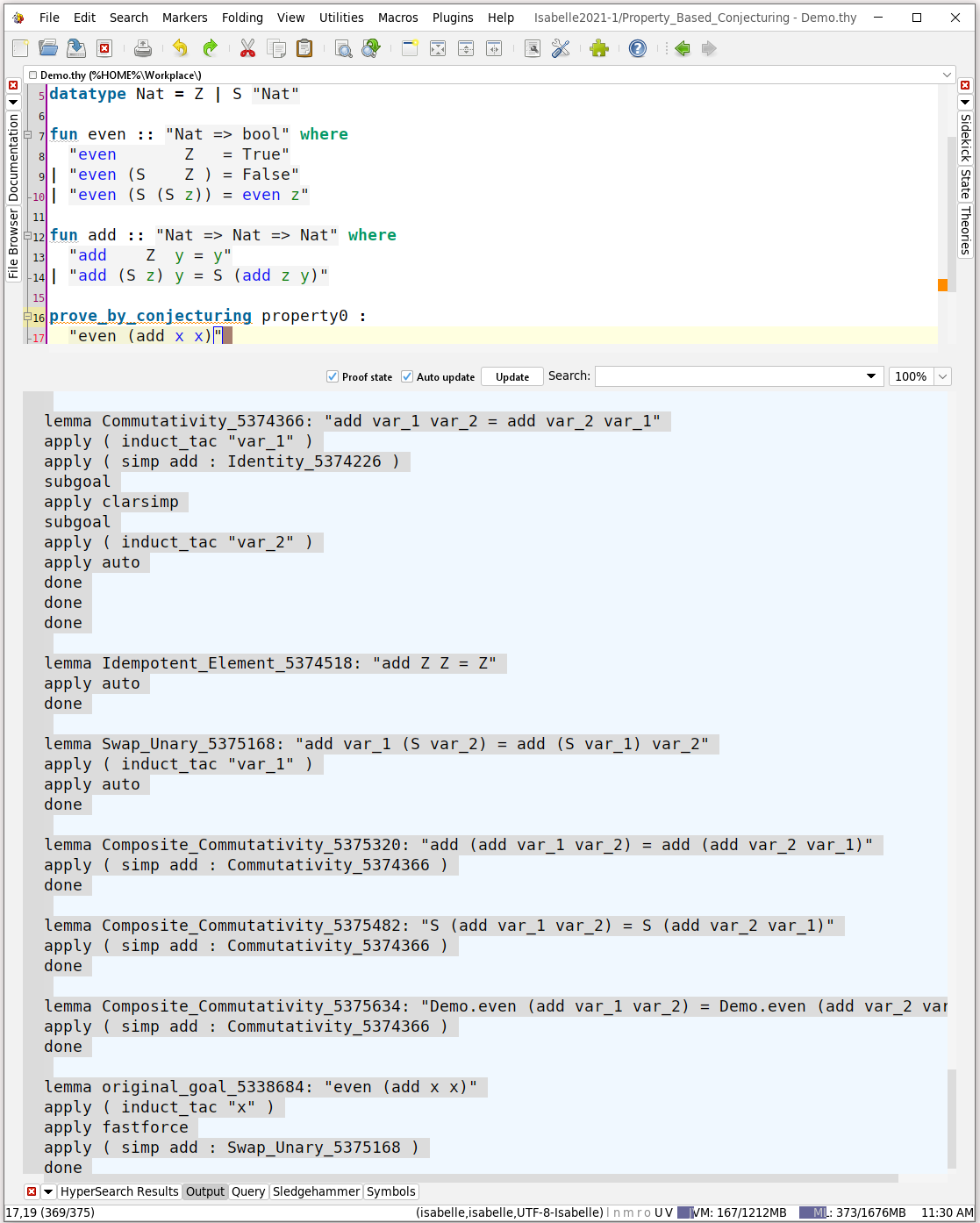}
\caption{Screenshot of Isabelle/HOL with TBC.
The upper pane shows the definition of a type and functions.
The new command \texttt{prove\_by\_conjecturing} invokes TBC, which presents the proof script appearing 
in the lower pane.} \label{figure:screenshot}
\end{figure}

The seamless integration of TBC into the Isabelle ecosystem
lets users build TBC as an Isabelle theory using Isabelle's standard build command without installing additional software.
Furthermore, when TBC finds a proof for the original goal,
our tool shows the final proof as well as proved conjectures 
with their proofs in the output pane as shown in Fig. \ref{figure:screenshot}.
Users can copy and paste them with a single click to the right location of their proof scripts.
The produced scripts are human readable, and Isabelle can check them without TBC.

\subsection{Template-Based Conjecturing}\label{sec:properties}

As mentioned in Section \ref{sec:overview}, our tool produces conjectures based on 16 templates specified in advance.
12 of them are either well-known algebraic properties, such as associative template, 
or relational properties, such as transitivity.
Note that we added the 4 highlighted templates based on the feedback from students who manually solved several benchmark problems.
None of these templates are specific to particular functions.

To produce conjectures for such templates, TBC first collects 
functions appearing in the original proof goal.
Then, it looks for the definitions of these functions and adds functions in 
these definitions into the list of functions for conjecturing.
Then, TBC filters out functions defined within the standard library
since the standard library already contains useful auxiliary lemmas for them. 
Finally, TBC creates conjectures by filling templates with these functions.

\begin{program}[t]
\begin{alltt}
associativity \hfill{} f (f (\(x\), \(y\)), \(z\)) = f (\(x\), f (\(y\), \(z\)))
identity element \hfill{} f (e, \(x\)) = \(x\) or f (\(x\), e) = \(x\) for some e
commutativity \hfill{} f (\(x\), \(y\)) = f (\(y\), \(x\))
idempotent element \hfill{} f (e, e) = e for some e
idempotency \hfill{} f (\(x\), \(x\)) = \(x\)
distributivity \hfill{} f (\(x\), g (\(y\), \(z\))) = g (f (\(x\), \(y\)), f (\(x\), \(z\)))
anti-distributivity \hfill{} f (g (\(x\), \(y\))) = g (f \(y\), f \(x\))
homomorphism \hfill{} f (g (\(x\), \(y\))) = g (f \(x\), f \(y\))
transitivity \hfill{} \(x\) R \(y\) \(\xrightarrow{}\) \(y\) R \(z\) \(\xrightarrow{}\) \(x\) R \(z\)
symmetry \hfill{} \(x\) R \(y\) \(\xrightarrow{}\) \(y\) R \(x\)
connexity \hfill{} \(x\) R \(y\) \(\lor\) \(y\) R \(x\) \(\lor\) (\(x\) = \(y\))
reflexivity \hfill{} \(x\) R \(x\)
\hl{square} \hfill{} f (f \(x\)) = \(x\)
\hl{swap-unary} \hfill{} f (\(x\), g \(y\)) = f (g \(x\), \(y\))
\hl{projection} \hfill{} f (f \(x\)) = f \(x\)
\hl{composite_commutativity} \hfill{} f (g (\(x\), \(y\))) = f (g (\(y\), \(x\)))
\end{alltt}
\caption{The Complete List of Template-Based Conjectures.
We added the highlighted four conjectures after manually solving some benchmark problems.
One can see that none of these conjectures are specific to particular problems.}
\label{code:full_properties}
\end{program}

\section{Evaluation}\label{sec:evaluation}

\subsection{Benchmark and Environment}

We evaluated our tool using Tons of Inductive Problems (TIP) \cite{tip},
which is a benchmark consisting of 462 inductive problems.
TIP consists of three main problem sets: 85 problems in Isaplanner, 50 in Prod, and 327 in TIP15.
Isaplanner is the easiest, whereas Prod contains problems at the intermediate difficulty level,
and TIP15 has difficult problems, such as Fermat's Last Theorem.

The advantage of using TIP is that each problem is complete within a single file.
That is, data types and functions are defined afresh within each problem file, 
instead of using the standard definition.
For example, our running example problem from Section \ref{introduction} 
is formalised as an independent Isabelle theory file in the Prod set in TIP.
The functions, \verb|add| and \verb|even|, are defined afresh in this file,
instead of using the default ones from the standard library.
This allowed us to ignore manually developed lemmas
for similar functions in the standard library.
This way, by using TIP, we focused on TBC's conjecturing capability to prove the final goal.

In this experiment, we set the following timeouts for the counter-example generators:
one second for Quickcheck, two seconds for Nitpick.
The timeout for Sledgehammer is more flexible:
10 seconds when attacking conjectures in $n$-th round where $n$ is an odd number,
whereas 30 seconds when attacking conjectures in $n$-th round where $n$ is an even number
or attacking the original goals.

However, when measuring the performance of TBC against TIP15 problems, we set the following short timeouts
to process 327 problems using computational resources available to us: 5 seconds for Sledgehammer to prove produced conjectures, 10 seconds for Sledgehammer to attack the original goal.
Furthermore, We use 15 minutes as the overall timeout for each problem in TIP15.

We ran our evaluations on consumer-grade laptops. 
Specifically, we used a \verb|Lenovo Thinkpad T490s|, with \verb|Intel Core i7-8665U| CPU and 16GB of RAM. We used \verb|Windows 10 Pro| as our evaluation operating system.

\subsection{Results}

\paragraph{Success rates for different difficulty levels.}
Figure \ref{fig:success_rates_problem_sets} shows 
the percentage of problems proved by each tool at each stage.
We use an induction prover for Isabelle/HOL \cite{definitional}, \texttt{TAP21},
as our baseline prover.
``Round0'' represents the percentage of solved problems after the zeroth round of TBC,
where \texttt{TBC} shows the percentage of solved problems after the second round
for Isaplanner and Prod, but after the first round for TIP15
due to our limited computational resources.

The figure shows that TBC brought the largest
improvement (40 percentage points) to the Prod category.
On the other hand, 
we can prove 60\% of problems in Isaplanner without producing conjectures,
while TBC struggles at harder problems in the TIP15 category.

\begin{program}[ht!]
\begin{alltt}
Ors [
  Auto_Solve,
  PThenOne [Old_Smart_Induct, Auto_Solve],
  PThenOne [Old_Smart_Induct,
    Thens [ Auto, RepeatN (Hammer), IsSolved ]
  ]
]
\end{alltt}
\caption{\texttt{TAP\_2021} is the strategy used in the baseline prover introduced by Nagashima \cite{definitional}. 
Since we added minor improvements to \texttt{Smart\_Induct}, we represent their version of \texttt{Smart\_Induct} as \texttt{Old\_Smart\_Induct} in this paper.}
\label{code:tap21_strategy}
\end{program}

\input{graph_overall}

\paragraph{Proof completion rates and execution time.}

Fig. \ref{fig:success_rate_time_isaplanner}, Fig. \ref{fig:success_rate_time}, and Fig. \ref{fig:success_rate_time_tip15} show the chances of solving a problem in each category relative to how long the program is run. 
For example, Fig. \ref{fig:success_rate_time} illustrates that approximately 20\% of the problems are solved within 5 minutes in the Prod category, and 60\% of the problems are solved within 20 minutes of runtime. 
Beyond this time, the chances of producing a proof increase marginally, reaching 66\% of problems after an hour. 

\input{graph_isaplanner_time}
\input{graph_prod_time}
\input{graph_tip15_time}

\paragraph{Refuting and proving.}
Fig. \ref{fig:conjectures_isaplanner} and Fig. \ref{fig:conjectures} show
how many conjectures TBC produced for each problem in Isaplanner and Prod
and how it handled them, respectively.
As shown in the figure, TBC did not produce any conjectures for some problems,
since it proved these problems even before producing conjectures.
Furthermore,
the number of conjectures does not blow up in TBC,
since TBC produces conjectures about commonly used properties only.
Note that keeping the number of conjectures low is the main challenge in other conjecturing tools, 
as we discuss in Section \ref{sec:related_work}.
Moreover,
these figures show that
most conjectures are either proved or refuted for problems in Isaplanner and Prod, 
and only a few conjectures are left unsolved
thanks to the strong default strategy and counter-example finders.

\input{graph_isaplanner_conjecture}
\input{graph_prod_conjecture}

\section{Related Work}\label{sec:related_work}

\paragraph{Conjecturing.}
We have two schools of conjecturing to automate inductive theorem proving:
top-down approaches and bottom-up approaches.
Top-down approaches \cite{rippling,oyster_clam,pgt} create auxiliary lemmas from an ongoing proof attempt,
whereas bottom-up approaches \cite{hipspec,hipster} produce lemmas from available functions and data types
to enrich the background theory \cite{moas_survey}.
TBC falls into the latter category.
While most bottom-up tools, such as HipSpec \cite{hipspec} and Hipster\cite{hipster},
produce conjectures randomly, TBC makes conjectures based on a fixed set of templates.
Furthermore, Hipster aims to \textit{discover} new lemmas, TBC checks for \textit{known} properties to keep the number of conjectures low.
In this respect,
RoughSpec \cite{template_based_conjecturing} is similar to TBC:
it produces conjectures based on templates, which 
describe important properties.
Contrary to TBC, RoughSpec supports only equations as templates and is a tool for Haskell
rather than a proof assistant.

\paragraph{Inductive theorem proving.}
TBC is an automatic tool developed for an interactive theorem prover (ITP)
based on a higher-order logic.
Others have introduced proof by induction for automatic theorem provers (ATPs)
\cite{acl2,induction_in_saturation,vampire_recursion,induction_for_smt,imandra}.
ATPs are typically based on less expressive logics
and use different proof calculi
compared to LCF-style provers.
Moreover, ATPs are built for performance,
whereas LCF-style provers are designed for high assurance and easy user-interaction.
Such differences make a straightforward comparison difficult;
however, we argue that
a stronger automation of inductive proofs in ITPs helps users reason data types and functions they introduce to tackle unique problems.

\section{Discussion and Conclusion}

Careful investigations into generated proofs reveal that
TBC proves conjectures that are not used to attack
the original goal as shown in Appendix.
Although such conjectures may serve as auxiliary lemmas
when users prove other problems in the future,
the time spent to prove these conjectures certainly slows
down the execution speed of TBC.
Furthermore, TBC fails to prove difficult problems
since they require conjectures specific to them.
We expect that combining TBC with other top-down approaches would 
result in more powerful automation, which remains as our future work.

This paper presented our template-based conjecturing tool, TBC.
To the best of our knowledge, TBC is the only tool that achieved high proof completion rates 
for the TIP benchmarks 
while producing human readable 
proofs that are native to a widely used ITP.

In this work we used 12 commonly known algebraic properties and 4 manually identified useful conjectures as our templates.
It remains our future work to incorporate templates that are found automatically by analysing large databases \cite{lol} into our framework.

\bibliographystyle{splncs04}
\bibliography{main.bib}

\newpage
\section*{Appendix}
\begin{program}[H]
\begin{alltt}
lemma associativity_5382114:
  "add var_1 (add var_2 var_3) = add (add var_1 var_2) var_3"
  apply ( induct "var_1" arbitrary : var_2 ) apply auto done
lemma associativity_5382286:
  "add (add var_1 var_2) var_3 = add var_1 (add var_2 var_3)"
  apply ( induct "var_1" arbitrary : var_2 ) apply auto done
lemma identity_5382450: "add 0 var_1 = var_1" apply auto done
lemma \hl{identity_5382590}: "add var_1 0 = var_1"
  apply ( induct_tac "var_1" ) apply auto done
lemma \hl{commutativity_5382730}:
  "add var_1 var_2 = add var_2 var_1" apply ( induct_tac "var_1" )
  apply ( simp add : \hl{identity_5382590} )
  subgoal apply clarsimp subgoal apply ( induct_tac "var_2" )
  apply auto done done done
lemma idempotent_Element_5382882: "add 0 0 = 0" apply auto done
lemma swap_Unary_5383532: "add var_1 (S var_2) = add (S var_1) var_2"
  apply ( induct_tac "var_1" ) apply auto done
lemma composite_Commutativity_5383684:
  "add (add var_1 var_2) = add (add var_2 var_1)"
  apply ( simp add : commutativity_5382730 ) done
lemma composite_Commutativity_5383846:
  "S (add var_1 var_2) = S (add var_2 var_1)"
  apply ( simp add : commutativity_5382730 ) done
lemma composite_Commutativity_5383998:
  "even (add var_1 var_2) = even (add var_2 var_1)"
  apply ( simp add : commutativity_5382730 ) done
lemma \hl{original_goal_5347090}: "even (add x x)" apply ( induct_tac "x" ) 
  apply fastforce apply ( metis Nat.distinct ( 1 ) Nat.inject 
  even.simps(3) \hl{commutativity_5382730} add.elims ) done
\end{alltt}
\caption{Generated Proof Script for Our Running Example}
\label{code:generated_proof}
\end{program}

Program \ref{code:generated_proof} shows the output of TBC
for our running example.
The original goal is proved using \texttt{commutativity\_5382730},
which is in turn proved using \texttt{identity\_5382590}.
8 out of 10 proved conjectures are not used to prove the final goal;
however, TBC outputs them, so that users may exploit them in future.

\end{document}

%% file: graph_overall.tex
\begin{figure}[ht!]
\centering
\begin{tikzpicture}[thick, scale=0.9]
\tikzstyle{every node}=[font=\large]
\begin{axis}[
    bar width = 14pt,
    ybar,
    enlargelimits=0.15,
    legend style={at={(0.5,-0.15)},
      anchor=north,legend columns=-1},
    ylabel={Proved problems [\%]},
    symbolic x coords={Isaplanner, Prod, TIP15},
    xtick=data,
    nodes near coords,
    nodes near coords align={vertical},
    ]
\addplot[pattern=crosshatch dots] coordinates {(Isaplanner, 53) (Prod, 16) (TIP15, 10)};
\addplot[] coordinates {(Isaplanner, 60) (Prod, 16) (TIP15, 12)};
\addplot[pattern=north east lines] coordinates {(Isaplanner, 75) (Prod, 66) (TIP15, 14)};
\legend{\texttt{TAP21} \ \ , Round0 \ \ , \texttt{PBC}}
\end{axis}
\end{tikzpicture}
\caption[first caption.]{Proof completion rates.}
\label{fig:success_rates_problem_sets}
\end{figure}
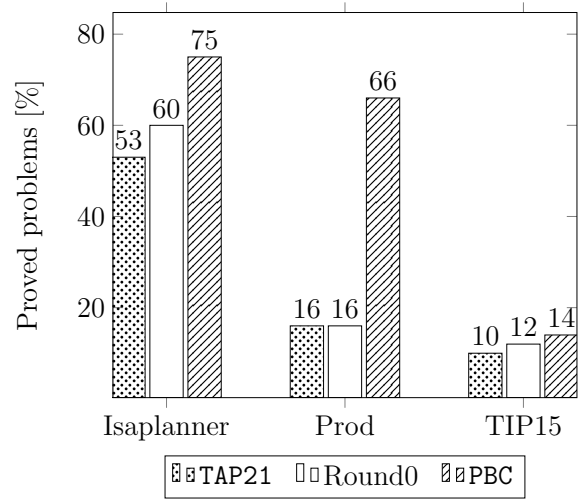

%% file: graph_isaplanner_time.tex
\begin{figure}[ht!]
\centering
\begin{tikzpicture}[thick, scale=0.9]
\tikzstyle{every node}=[font=\large]
\begin{axis}[
    title={},
    xlabel={Time [minute]},
    ylabel={Chances of solving [\%]},
    xmin=0, xmax=60,
    ymin=0, ymax=80,
    xtick={0, 10, 20, 30, 40, 50, 60},
    ytick={0, 20, 40, 60, 80, 100},
    ymajorgrids=true,
    grid style=dashed,
    legend pos=south east,
]
\addplot[color=blue, dashed] coordinates {(0.0, 1.1764705882352942)(0.0, 2.3529411764705883)(0.0, 3.5294117647058822)(0.0, 4.705882352941177)(0.0, 5.882352941176471)(0.0, 7.0588235294117645)(0.0, 8.235294117647058)(0.0, 9.411764705882353)(0.0, 10.588235294117647)(0.0, 11.764705882352942)(0.0, 12.941176470588236)(0.0, 14.117647058823529)(0.0, 15.294117647058824)(0.0, 16.470588235294116)(0.0, 17.647058823529413)(0.0, 18.823529411764707)(0.01, 20.0)(0.01, 21.176470588235293)(0.01, 22.352941176470587)(0.01, 23.529411764705884)(0.01, 24.705882352941178)(0.01, 25.88235294117647)(0.01, 27.058823529411764)(0.01, 28.235294117647058)(0.01, 29.41176470588235)(0.01, 30.58823529411765)(0.01, 31.764705882352942)(0.01, 32.94117647058823)(0.01, 34.11764705882353)(0.01, 35.294117647058826)(0.02, 36.470588235294116)(0.02, 37.64705882352941)(0.02, 38.8235294117647)(0.02, 40.0)(0.02, 41.1764705882353)(0.03, 42.35294117647059)(0.03, 43.529411764705884)(0.69, 44.705882352941174)(0.7, 45.88235294117647)(0.72, 47.05882352941177)(2.11, 48.23529411764706)(2.6, 49.411764705882355)(2.98, 50.588235294117645)(3.19, 51.76470588235294)(3.89, 52.94117647058823)(60.0, 52.94117647058823)};
\addlegendentry{TAP21};

\addplot[color=blue, ] coordinates {(0.0, 1.1764705882352942)(0.0, 2.3529411764705883)(0.0, 3.5294117647058822)(0.0, 4.705882352941177)(0.0, 5.882352941176471)(0.0, 7.0588235294117645)(0.0, 8.235294117647058)(0.0, 9.411764705882353)(0.0, 10.588235294117647)(0.0, 11.764705882352942)(0.01, 12.941176470588236)(0.01, 14.117647058823529)(0.01, 15.294117647058824)(0.01, 16.470588235294116)(0.01, 17.647058823529413)(0.01, 18.823529411764707)(0.01, 20.0)(0.01, 21.176470588235293)(0.01, 22.352941176470587)(0.01, 23.529411764705884)(0.01, 24.705882352941178)(0.01, 25.88235294117647)(0.01, 27.058823529411764)(0.01, 28.235294117647058)(0.01, 29.41176470588235)(0.01, 30.58823529411765)(0.01, 31.764705882352942)(0.02, 32.94117647058823)(0.02, 34.11764705882353)(0.02, 35.294117647058826)(0.02, 36.470588235294116)(0.02, 37.64705882352941)(0.02, 38.8235294117647)(0.02, 40.0)(0.03, 41.1764705882353)(0.03, 42.35294117647059)(0.37, 43.529411764705884)(2.16, 44.705882352941174)(3.15, 45.88235294117647)(3.34, 47.05882352941177)(3.59, 48.23529411764706)(3.59, 49.411764705882355)(3.67, 50.588235294117645)(3.9, 51.76470588235294)(3.91, 52.94117647058823)(4.05, 54.11764705882353)(4.23, 55.294117647058826)(4.67, 56.470588235294116)(4.79, 57.64705882352941)(5.12, 58.8235294117647)(5.14, 60.0)(5.46, 61.1764705882353)(5.93, 62.35294117647059)(11.53, 63.529411764705884)(11.94, 64.70588235294117)(15.22, 65.88235294117646)(17.63, 67.05882352941177)(21.19, 68.23529411764706)(21.88, 69.41176470588235)(24.81, 70.58823529411765)(27.95, 71.76470588235294)(29.61, 72.94117647058823)(48.89, 74.11764705882354)(50.12, 75.29411764705883)(60.0, 75.29411764705883)};
\addlegendentry{PBC};

\end{axis}
\end{tikzpicture}
\caption[first caption.]{Success rates over time for Isaplanner.}\label{fig:success_rate_time_isaplanner}
\end{figure}
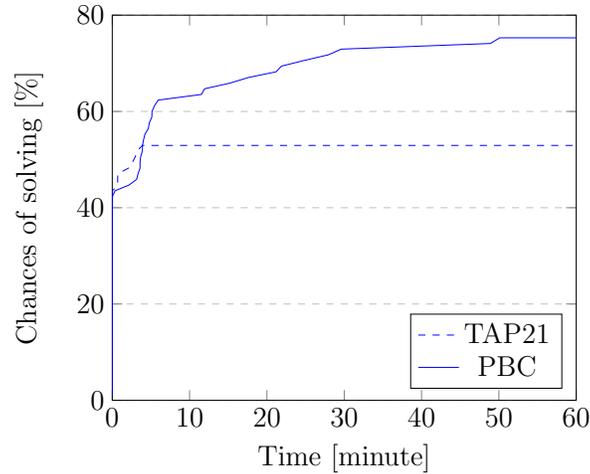

%% file: graph_prod_time.tex
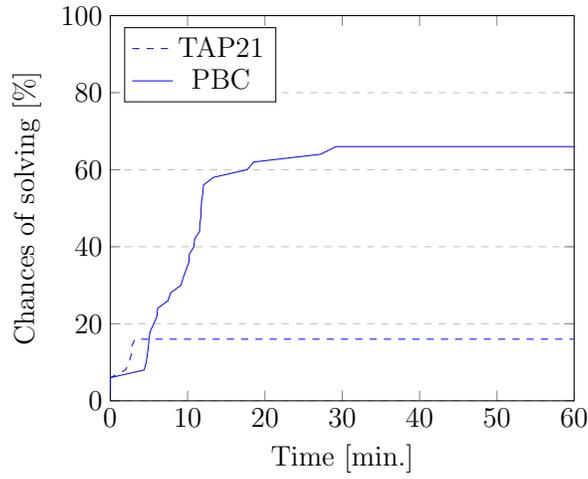
\begin{figure}[ht!]
\centering
\begin{tikzpicture}[thick, scale=0.9]
\tikzstyle{every node}=[font=\large]
\begin{axis}[
    title={},
    xlabel={Time [min.]},
    ylabel={Chances of solving [\%]},
    xmin=0, xmax=60,
    ymin=0, ymax=100,
    xtick={0, 10, 20, 30, 40, 50, 60},
    ytick={0, 20, 40, 60, 80, 100},
    ymajorgrids=true,
    grid style=dashed,
    legend pos=north west
]
\addplot[color=blue, dashed] coordinates {(0.01, 2.0)(0.01, 4.0)(0.01, 6.0)(1.97, 8.0)(2.43, 10.0)(2.72, 12.0)(2.73, 14.0)(3.19, 16.0)(60.0, 16.0)};
\addlegendentry{TAP21};

\addplot[color=blue, ] coordinates {(0.01, 2.0)(0.01, 4.0)(0.02, 6.0)(4.38, 8.0)(4.66, 10.0)(4.79, 12.0)(4.93, 14.0)(5.01, 16.0)(5.16, 18.0)(5.63, 20.0)(6.05, 22.0)(6.12, 24.0)(7.48, 26.0)(7.8, 28.0)(9.14, 30.0)(9.44, 32.0)(9.8, 34.0)(10.18, 36.0)(10.21, 38.0)(10.82, 40.0)(10.91, 42.0)(11.55, 44.0)(11.59, 46.0)(11.74, 48.0)(11.75, 50.0)(11.82, 52.0)(11.97, 54.0)(12.05, 56.0)(13.32, 58.0)(17.71, 60.0)(18.54, 62.0)(27.19, 64.0)(29.2, 66.0)(60.0, 66.0)};
\addlegendentry{PBC};

\end{axis}
\end{tikzpicture}
\caption[first caption.]{Success rates over time for Prod.}\label{fig:success_rate_time}

\end{figure}

%% file: graph_tip15_time.tex
\begin{figure}[ht!]
\centering
\begin{tikzpicture}[thick, scale=0.9]
\tikzstyle{every node}=[font=\large]
\begin{axis}[
    title={},
    xlabel={Time [minute]},
    ylabel={Chances of solving [\%]},
    xmin=0, xmax=15,
    ymin=0, ymax=15,
    xtick={0, 5, 10, 15, 20, 30, 40, 50, 60},
    ytick={0, 5, 10, 15, 20, 40, 60, 80, 100},
    ymajorgrids=true,
    grid style=dashed,
    legend pos=south east,
]
\addplot[color=blue, dashed] coordinates {(0.0, 0.3058103975535168)(0.0, 0.6116207951070336)(0.0, 0.9174311926605505)(0.0, 1.2232415902140672)(0.0, 1.529051987767584)(0.0, 1.834862385321101)(0.0, 2.140672782874618)(0.0, 2.4464831804281344)(0.0, 2.7522935779816513)(0.0, 3.058103975535168)(0.0, 3.363914373088685)(0.0, 3.669724770642202)(0.0, 3.9755351681957185)(0.0, 4.281345565749236)(0.0, 4.587155963302752)(0.0, 4.892966360856269)(0.01, 5.198776758409786)(0.01, 5.504587155963303)(0.01, 5.81039755351682)(0.01, 6.116207951070336)(0.01, 6.422018348623853)(0.01, 6.72782874617737)(0.02, 7.033639143730887)(0.09, 7.339449541284404)(0.71, 7.6452599388379205)(0.93, 7.951070336391437)(1.14, 8.256880733944953)(1.38, 8.562691131498472)(1.42, 8.868501529051988)(1.74, 9.174311926605505)(1.9, 9.480122324159021)(1.97, 9.785932721712538)(2.61, 10.091743119266056)(2.89, 10.397553516819572)(60.0, 10.759493670886076)};
\addlegendentry{TAP21};

\addplot[color=blue, ] coordinates {(0.0, 0.3058103975535168)(0.0, 0.6116207951070336)(0.0, 0.9174311926605505)(0.0, 1.2232415902140672)(0.0, 1.529051987767584)(0.0, 1.834862385321101)(0.0, 2.140672782874618)(0.0, 2.4464831804281344)(0.0, 2.7522935779816513)(0.0, 3.058103975535168)(0.0, 3.363914373088685)(0.0, 3.669724770642202)(0.0, 3.9755351681957185)(0.01, 4.281345565749236)(0.01, 4.587155963302752)(0.01, 4.892966360856269)(0.01, 5.198776758409786)(0.01, 5.504587155963303)(0.01, 5.81039755351682)(0.01, 6.116207951070336)(0.02, 6.422018348623853)(0.09, 6.72782874617737)(1.5, 7.033639143730887)(2.4, 7.339449541284404)(2.41, 7.6452599388379205)(2.9, 7.951070336391437)(2.93, 8.256880733944953)(2.99, 8.562691131498472)(2.99, 8.868501529051988)(2.99, 9.174311926605505)(3.06, 9.480122324159021)(3.16, 9.785932721712538)(3.38, 10.091743119266056)(3.64, 10.397553516819572)(3.73, 10.703363914373089)(4.03, 11.009174311926605)(4.47, 11.314984709480122)(4.92, 11.62079510703364)(4.93, 11.926605504587156)(5.06, 12.232415902140673)(5.14, 12.53822629969419)(5.44, 12.844036697247706)(7.03, 13.149847094801224)(8.62, 13.45565749235474)(11.69, 13.761467889908257)(60.0, 14.240506329113924)};
\addlegendentry{PBC};

\end{axis}
\end{tikzpicture}
\caption[first caption.]{Success rates over time for TIP15.}\label{fig:success_rate_time_tip15}
\end{figure}
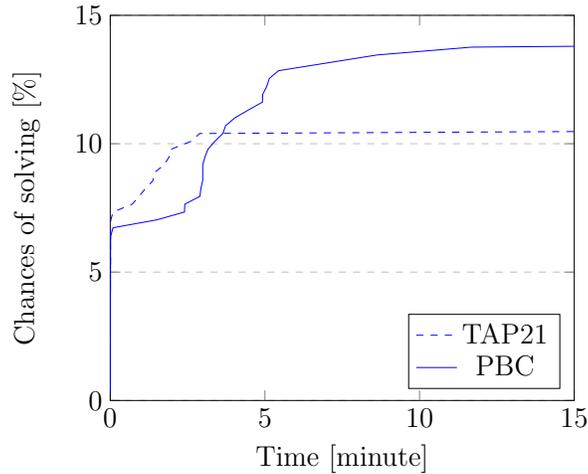

%% file: graph_isaplanner_conjecture.tex
\begin{figure}[ht!]

\centering
\begin{tikzpicture}[thick, scale=1.0]
\tikzstyle{every node}=[font=\large]
\begin{axis}[
    ybar stacked,
	bar width=2pt,
    enlargelimits=0.05,
    legend style={at={(0.5,-0.20)},
      anchor=north,legend columns=-1},
    ylabel={\# of conjectures},
    symbolic x coords={id0, id1, id2, id3, id4, id5, id6, id7, id8, id9, id10, id11, id12, id13, id14, id15, id16, id17, id18, id19, id20, id21, id22, id23, id24, id25, id26, id27, id28, id29, id30, id31, id32, id33, id34, id35, id36, id37, id38, id39, id40, id41, id42, id43, id44, id45, id46, id47, id48, id49, id50, id51, id52, id53, id54, id55, id56, id57, id58, id59, id60, id61, id62, id63, id63, id64, id65, id66, id67,
    id68, id69, id70, id71, id72, id73, id74, id75, id76, id77,
    id78, id79, id80, id81, id82, id83, id84, id85, id86},
    x tick label style={rotate=45,anchor=east},
    ]
\addplot+[pattern=north east lines,ybar] plot coordinates {
(id0, 0) (id1, 0) (id2, 0) (id3, 0) (id4, 0) (id5, 0) (id6, 0) (id7, 0) (id8, 0) (id9, 0) (id10, 0) (id11, 0) (id12, 0) (id13, 0) (id14, 0) (id15, 0) (id16, 0) (id17, 0) (id18, 0) (id19, 3) (id20, 0) (id21, 0) (id22, 8) (id23, 8) (id24, 0) (id25, 0) (id26, 0) (id27, 0) (id28, 6) (id29, 0) (id30, 0) (id31, 0) (id32, 0) (id33, 0) (id34, 0) (id35, 0) (id36, 0) (id37, 0) (id38, 0) (id39, 0) (id40, 0) (id41, 0) (id42, 0) (id43, 4) (id44, 11) (id45, 30) (id46, 17) (id47, 0) (id48, 13) (id49, 5) (id50, 24) (id51, 0) (id52, 7) (id53, 16) (id54, 7) (id55, 0) (id56, 0) (id57, 0) (id58, 0) (id59, 6) (id60, 0) (id61, 11) (id62, 5) (id63, 16) (id64, 7) (id65, 10) (id66, 0) (id67, 0) (id68, 25) (id69, 9) (id70, 25) (id71, 0) (id72, 0) (id73, 6) (id74, 7) (id75, 14) (id76, 0) (id77, 7) (id78, 7) (id79, 13) (id80, 13) (id81, 19) (id82, 8) };
\addplot+[pattern=mynewdots,ybar] plot coordinates {
(id0, 0) (id1, 0) (id2, 0) (id3, 0) (id4, 0) (id5, 0) (id6, 0) (id7, 0) (id8, 0) (id9, 0) (id10, 0) (id11, 0) (id12, 0) (id13, 0) (id14, 0) (id15, 0) (id16, 0) (id17, 0) (id18, 0) (id19, 0) (id20, 0) (id21, 0) (id22, 17) (id23, 17) (id24, 0) (id25, 0) (id26, 0) (id27, 0) (id28, 5) (id29, 0) (id30, 0) (id31, 0) (id32, 0) (id33, 0) (id34, 0) (id35, 0) (id36, 0) (id37, 0) (id38, 0) (id39, 0) (id40, 0) (id41, 0) (id42, 0) (id43, 10) (id44, 5) (id45, 6) (id46, 2) (id47, 0) (id48, 11) (id49, 4) (id50, 12) (id51, 0) (id52, 10) (id53, 2) (id54, 0) (id55, 0) (id56, 0) (id57, 0) (id58, 0) (id59, 2) (id60, 0) (id61, 12) (id62, 3) (id63, 2) (id64, 7) (id65, 13) (id66, 0) (id67, 0) (id68, 10) (id69, 7) (id70, 10) (id71, 0) (id72, 0) (id73, 12) (id74, 12) (id75, 2) (id76, 0) (id77, 11) (id78, 0) (id79, 6) (id80, 6) (id81, 8) (id82, 6) };
\addplot+[ybar] plot coordinates {
(id0, 0) (id1, 0) (id2, 0) (id3, 0) (id4, 0) (id5, 0) (id6, 0) (id7, 0) (id8, 0) (id9, 0) (id10, 0) (id11, 0) (id12, 0) (id13, 0) (id14, 0) (id15, 0) (id16, 0) (id17, 0) (id18, 0) (id19, 1) (id20, 0) (id21, 0) (id22, 5) (id23, 5) (id24, 0) (id25, 0) (id26, 0) (id27, 0) (id28, 1) (id29, 0) (id30, 0) (id31, 0) (id32, 0) (id33, 0) (id34, 0) (id35, 0) (id36, 0) (id37, 0) (id38, 0) (id39, 0) (id40, 0) (id41, 0) (id42, 0) (id43, 4) (id44, 0) (id45, 0) (id46, 0) (id47, 0) (id48, 0) (id49, 1) (id50, 0) (id51, 0) (id52, 0) (id53, 0) (id54, 0) (id55, 0) (id56, 0) (id57, 0) (id58, 0) (id59, 0) (id60, 0) (id61, 0) (id62, 0) (id63, 0) (id64, 0) (id65, 0) (id66, 0) (id67, 0) (id68, 0) (id69, 0) (id70, 0) (id71, 0) (id72, 0) (id73, 0) (id74, 1) (id75, 0) (id76, 0) (id77, 0) (id78, 0) (id79, 0) (id80, 0) (id81, 0) (id82, 0) };

\legend{\strut refuted \ \ , \strut proved \ \ , \strut not proved}
\end{axis}
\end{tikzpicture}
\caption{Conjectures for Isaplanner.}
\label{fig:conjectures_isaplanner}
\end{figure}
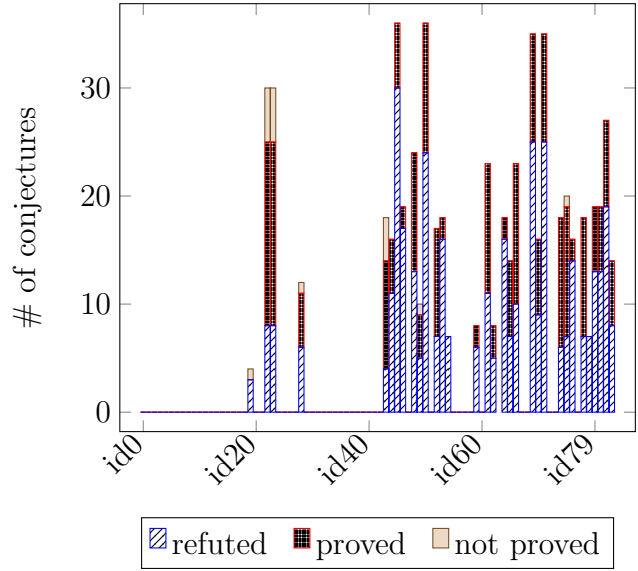

%% file: graph_prod_conjecture.tex
\begin{figure}[ht!]
\centering
\begin{tikzpicture}[thick, scale=1.0]
\tikzstyle{every node}=[font=\large]
\begin{axis}[
    ybar stacked,
	bar width=2pt,
    enlargelimits=0.05,
    legend style={at={(0.5,-0.20)},
      anchor=north,legend columns=-1},
    ylabel={\# of conjectures},
    symbolic x coords={id0, id1, id2, id3, id4, id5, id6, id7, id8, id9, id10, id11, id12, id13, id14, id15, id16, id17, id18, id19, id20, id21, id22, id23, id24, id25, id26, id27, id28, id29, id30, id31, id32, id33, id34, id35, id36, id37, id38, id39, id40, id41, id42, id43, id44, id45, id46, id47, id48,id49,id50},
    x tick label style={rotate=45,anchor=east},
    ]
\addplot+[pattern=north east lines,ybar] plot coordinates {
(id0, 10) (id1, 0) (id2, 0) (id3, 9) (id4, 11) (id5, 16) (id6, 15) (id7, 2) (id8, 2) (id9, 9) (id10, 9) (id11, 27) (id12, 12) (id13, 7) (id14, 7) (id15, 7) (id16, 9) (id17, 9) (id18, 9) (id19, 7) (id20, 7) (id21, 7) (id22, 9) (id23, 7) (id24, 12) (id25, 12) (id26, 27) (id27, 10) (id28, 27) (id29, 9) (id30, 8) (id31, 7) (id32, 0) (id33, 0) (id34, 0) (id35, 0) (id36, 0) (id37, 0) (id38, 0) (id39, 8) (id40, 8) (id41, 0) (id42, 3) (id43, 5) (id44, 5) };
\addplot+[pattern=mynewdots,ybar] plot coordinates {
(id0, 12) (id1, 0) (id2, 0) (id3, 6) (id4, 7) (id5, 17) (id6, 11) (id7, 0) (id8, 0) (id9, 6) (id10, 7) (id11, 8) (id12, 10) (id13, 12) (id14, 9) (id15, 10) (id16, 6) (id17, 6) (id18, 6) (id19, 6) (id20, 6) (id21, 6) (id22, 6) (id23, 10) (id24, 16) (id25, 10) (id26, 9) (id27, 8) (id28, 9) (id29, 6) (id30, 2) (id31, 6) (id32, 0) (id33, 0) (id34, 0) (id35, 0) (id36, 0) (id37, 0) (id38, 0) (id39, 16) (id40, 16) (id41, 0) (id42, 0) (id43, 13) (id44, 3) };
\addplot+[ybar] plot coordinates {
(id0, 0) (id1, 0) (id2, 0) (id3, 0) (id4, 1) (id5, 0) (id6, 1) (id7, 0) (id8, 0) (id9, 1) (id10, 0) (id11, 1) (id12, 0) (id13, 1) (id14, 0) (id15, 0) (id16, 1) (id17, 1) (id18, 1) (id19, 0) (id20, 0) (id21, 0) (id22, 0) (id23, 0) (id24, 0) (id25, 0) (id26, 0) (id27, 0) (id28, 0) (id29, 1) (id30, 0) (id31, 0) (id32, 0) (id33, 0) (id34, 0) (id35, 0) (id36, 0) (id37, 0) (id38, 0) (id39, 1) (id40, 1) (id41, 0) (id42, 1) (id43, 2) (id44, 2) };

\legend{\strut refuted \ \ , \strut proved \ \ , \strut not proved}
\end{axis}
\end{tikzpicture}
\caption{Conjectures for Prod.}
\label{fig:conjectures}
\end{figure}
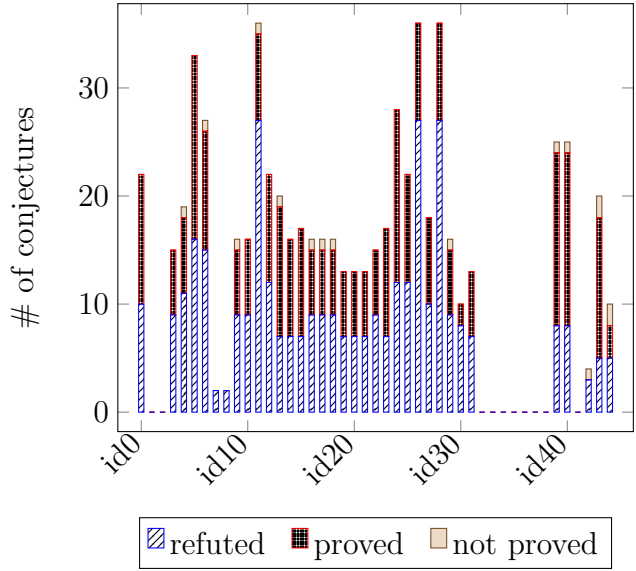